\documentclass[aps, prb, showpacs,
twocolumn, %showkeys, preprintnumbers,
amssymb,superscriptaddress]{revtex4}
\usepackage{graphicx, bm, amsmath, amsfonts,amssymb}

\usepackage{subfigure}

\def\v#1{\textbf{\emph{#1}}}

\def\dd{\mathrm{d}}

\begin{document}

%\preprint{}

\title{Time reversal symmetry breaking superconducting ground state in the
 doped Mott-insulator on honeycomb lattice}

\author{Zheng-Cheng Gu} \affiliation{Kavli
  Institute for Theoretical Physics, University of California, Santa
  Barbara, CA 93106, USA}

\author{Hong-Chen Jiang}
 \affiliation{Kavli
  Institute for Theoretical Physics, University of California, Santa
  Barbara, CA 93106, USA}

\author{D. N. Sheng}
\affiliation{Department of Physics and Astronomy, California State
University, Northridge, California 91330,USA}

\author{Hong Yao}
\affiliation{Department of Physics, Stanford University, Stanford,
CA 94305, USA}

\author{Leon Balents}
\affiliation{Kavli Institute for Theoretical Physics, University of
California, Santa Barbara, CA 93106, USA}

\author{Xiao-Gang Wen}
\affiliation{Department of Physics, Massachusetts Institute of Technology, Cambridge,
Massachusetts 02139, USA}

\date{\today}
\begin{abstract}
  The emergence of superconductivity in doped Mott insulators has been
  debated for decades.  In this paper, we report the theoretical
  discovery of a novel time reversal symmetry breaking superconducting
  ground state in the doped Mott-insulator(described by the well known $t$-$J$ model) on honeycomb lattice,
  based on a recently developed variational method:  the Grassmann
  tensor product state(GTPS) approach. As a benchmark, we use exact
  diagonalization (ED) and density matrix renormalization (DMRG)
  methods to check our results on small clusters. We find
  systematic consistency for the ground state energy as well as
  other physical quantities, such as the staggered magnetization. At
  low doping, the superconductivity coexists with anti-ferromagnetic
  ordering.
\end{abstract}


\maketitle

\section{Introduction}
Since the discovery of high-temperature superconductivity in
cuprates\cite{highTc}, many strongly correlated models have been
intensively studied. One of the simplest of these models is the $t$-$J$
model\cite{tJ}, which describes a doped Mott-insulator:
\begin{equation}
H_{t-J}=t\sum_{\langle ij\rangle,\sigma
}\tilde{c}^\dagger_{i,\sigma}
\tilde{c}_{j,\sigma}+h.c.+J\sum_{\langle ij \rangle}\left(\v S_i
\cdot\v S_j-\frac{1}{4}n_in_j\right),\label{tJ}
\end{equation}
where $\tilde{c}^\dagger_{i,\sigma}$ is the electron operator defined in
the no-double-occupancy subspace. This model can be derived from the
strong-coupling limit of the Hubbard model. It is believed that such a simple
model potentially captures the key mechanism of high $T_c$ cuprates. Despite its simplicity and
extensive study on it, the nature of the ground states of Eq.~\eqref{tJ} is
still controversial due to the no-double-occupancy constraint.

A strong correlation view of the $t$-$J$ model was advanced by
Anderson, who conjectured the relevance of a resonating valence bond
(RVB) state\cite{RVB} as a low energy state for Eq.~\eqref{tJ} when
doped.  When undoped, the RVB state is a spin singlet, with no
symmetry breaking, and describes a ``quantum spin liquid''. At low
temperature, the mobile carriers in the doped RVB state behave as
bosons and condense, forming a state indistinguishable in terms of
symmetry from a singlet BCS superconductor.  A further development
was the introduction of a projected mean-field wavefunction -- the
projection removing all components of the wavefunction with doubly
occupied sites -- which could be used
variationally\cite{d-wave,PatrickRVB}.

Presently, this variational method remains one of the few numerical
tools for $t$-$J$ like models which work directly at $T=0$ and
can deal with significant system sizes. However, due to the special form
of the variational wavefunction, one may be concerned about bias:
very general states in the low energy subspace cannot be
investigated.

Recently new numerical tools have been developed to investigate much
more general low energy states in $t$-$J$ like models beyond the
projective method. One novel construction builds Tensor Product
States
(TPS's)\cite{FrankPEPS,infinitPEPS,GuTERG,XiangTRG1,XiangTRG2,XiangTRG3},
which can be conveniently studied and have been applied to many spin
systems\cite{infinitPEPS,GuTERG,XiangTRG1,XiangTRG2,XiangTRG3,TPSspin,TPSfspin}.
This new class of variational states do not assume any specifical
ordering pattern a priori and can describe very general states as
long as their entanglement entropies satisfy perimeter law.
Recently, this method has been generalized to fermionic
systems\cite{FrankfPEPS,ifPEPS,fermionicTPS,GPEPS,finitefPEPS,GuGTPS}.
Among many different generalizations, the Grassmann tensor product
states (GTPS's)\cite{GuGTPS} were shown to be closely related to
projective states.  They are able to describe a class of projective
wavefunctions faithfully, including the short-range
RVB states in particular . Very recently, the application of this kind of new numerical method
to the $t$-$J$ model on square lattice has reported\cite{squaretJ} the discovery of a stripe state
instead of the d-wave superconductivity suggested by the meanfield approach or projective
wave function approach a long time ago.

%Science the TPS and GTPS construction is particularly well suited to
%trivalent lattices, the simplest of which is the honeycomb lattice
%in two dimensions.
Since the ground state of square lattice $t$-$J$ model is still controversial, it is interesting to
investigate the phase diagram of the $t$-$J$ model on another lattice geometry, e.g., the honeycomb lattice. Like the square lattice appropriate for the
cuprates, the honeycomb lattice is bipartite and naturally supports
an antiferromagnetic (AF) state at half-filling in the strong
coupling (Heisenberg) limit.  Similarities to cuprate physics may be
expected. Moreover, several numerical studies have identified a
possible quantum spin liquid state on this lattice at half-filling
when additional quantum fluctuations are included in the
Hubbard\cite{honeycombSL} and Heisenberg\cite{EDhoneycomb,SLJ1J2}
models. Thus the doped $t$-$J$ model on the honeycomb lattice seems
a promising venue to explore RVB ideas.

Here, we investigate the ground state of this system using the
recently developed GTPS approach.  The GTPS results are benchmarked
by comparison with exact diagonalization (ED) and density matrix
renormalization group (DMRG) calculations.  Our results are
systematically consistent with these non-variational, exact methods
on small clusters.  The principle result of the GTPS calculations is
that the ground state at non-zero doping is a time reversal symmetry
breaking $d+id$ wave superconductor.  Some physical rationales for
this result are given at the end of this paper.

\begin{figure}[h]
{\includegraphics[width=0.3\textwidth]{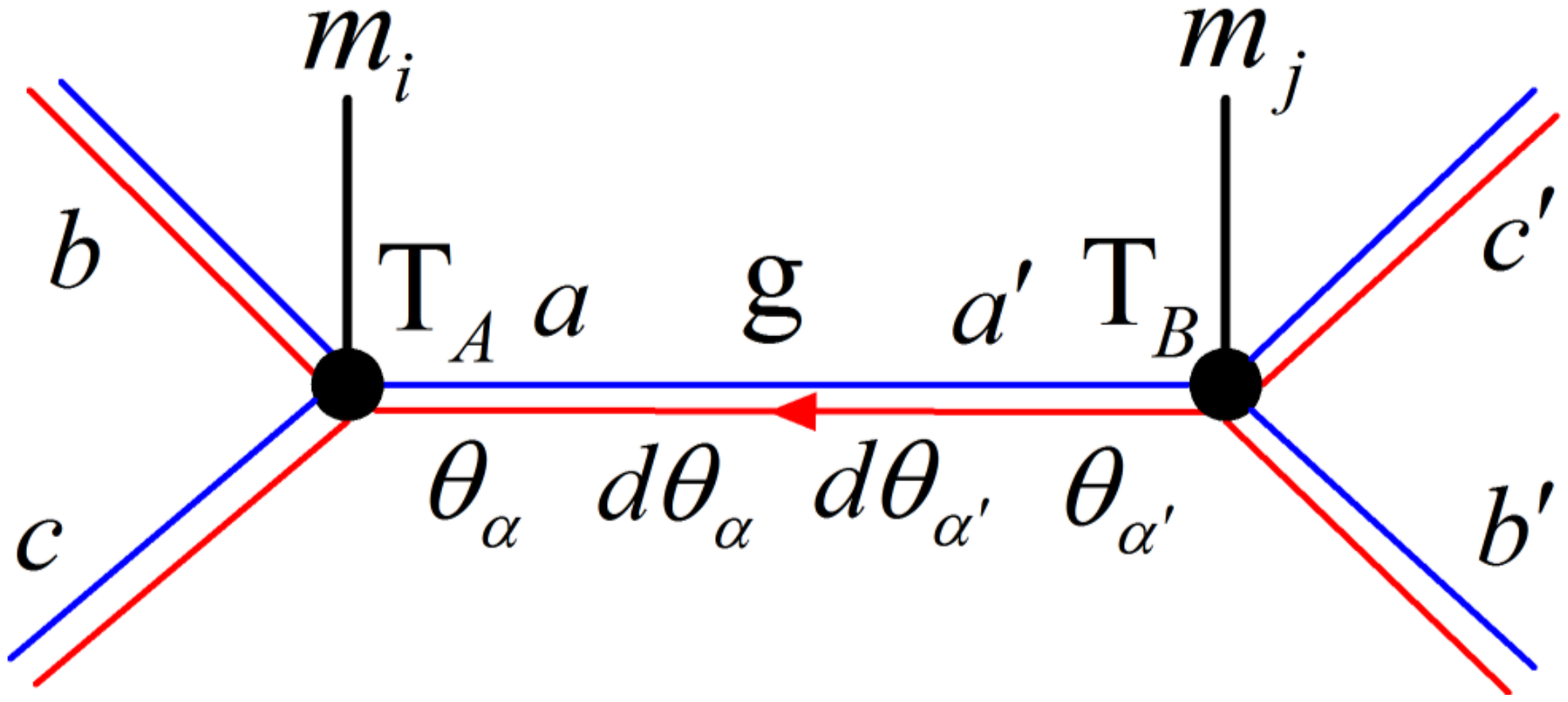}}
\caption{(Color online)Graphic representation of the GTPS on a honeycomb lattice.
$\textbf{T}_A$ and $\textbf{T}_B$, which contain $\theta$ are
defined on the sublattices $A$ and $B$ for each unit cell. The
Grassmann metric $\textbf{g}$ containing $\dd \theta$ is defined on
the links that connect the Grassmann tensors $\textbf{T}_A$ and
$\textbf{T}_B$. The blue lines represent the fermion parity even
indices while the red lines represent the fermion parity odd indices
of the virtual states. Notice that an arrow from $A$ to $B$ represents
the ordering convention $\dd \theta_\alpha\dd
\theta_{\alpha^\prime}$ that we use for the Grassmann
metric.}\label{fig:GrassmannTPS}
\end{figure}

\section{The variational ansatz}
We use the standard form of GTPS
as our variational wavefunction. We further assume a translationally
invariant ansatz, and thus it is specified by just two different
Grassmann tensors $\textbf{T}_A,\textbf{T}_B$ on sublattice $A,B$ of
each unit cell:
\begin{eqnarray}
&& \Psi(\{m_i\},\{m_j \}) \\
\nonumber &&  = {\rm{tTr}} \int \prod_{\langle ij\rangle}\textbf{g}_{a
a^\prime} \prod_{i\in A} {\textbf{T}}^{m_i}_{A;abc} \prod_{j\in B}
{\textbf{T}}^{m_j}_{B;a^\prime b^\prime c^\prime},\label{GTPS}
\end{eqnarray}
with
\begin{eqnarray}
{\textbf{T}}^{m_{i}}_{A;abc}&=& {T}^{m_{i}}_{A;abc}
\theta_\alpha^{P^f(a)} \theta_\beta^{P^f(b)}
\theta_\gamma^{P^f(c)},  \nonumber\\
 {\textbf{T}}^{m_{j}}_{B,a^\prime b^\prime c^\prime}&=&
{T}^{m_{j}}_{B;a^\prime b^\prime c^\prime}
\theta_{\alpha^\prime}^{P^f(a^\prime)}
\theta_{\beta^\prime}^{P^f(b^\prime)}
\theta_{\gamma^\prime}^{P^f(c^\prime)},    \nonumber\\
\textbf{g}_{aa^\prime}&=& \delta_{aa^\prime}{\dd
\theta}_\alpha^{P^f(a)} {\dd
\theta}_{\alpha^\prime}^{P^f(a^\prime)}.
\end{eqnarray}
We notice that the symbol $\rm{tTr}$ means tensor contraction of the
inner indices $\{a\}$. Here
$\theta_{\alpha(\beta,\gamma)},\mathrm{d}\theta_{\alpha(\beta,\gamma)}$
are the Grassmann numbers and dual Grassmann numbers respectively
defined on the link $a(b,c)$and they satisfy the Grassmann algebra:
\begin{align}
 \theta_\alpha\theta_\beta&=-\theta_\beta\theta_\alpha,
&
 \dd{\theta_\alpha}\dd{\theta_\beta}&=-\dd{\theta_\beta}\dd{\theta_\alpha},
\nonumber\\
\int \dd{\theta_\alpha}\theta_\beta &=\delta_{\alpha\beta} & \int
\dd{\theta_\alpha} 1&=0 .
\end{align}
As shown in Fig.\ref{fig:GrassmannTPS}, $a,b,c=1,2,\ldots,D$ are the
virtual indices carrying a fermion parity $P^f(a)=0,1$. In this
paper, we choose $D$ to be even and assume that there are \emph{equal}
numbers of fermion parity even/odd indices, which might be not
necessary in general. Those indices with odd parity are always
associated with a Grassmann number on the corresponding link and the
metric $\textbf{g}_{aa^\prime}$ is the Grassmann generalization of
the canonical delta function. The complex coefficients
${T}^{m_{i}}_{A;abc}$ and ${T}^{m_{j}}_{B;a^\prime b^\prime
c^\prime}$ are the variational parameters.

Notice that $m_{i}$ is the physical index of the $t$-$J$ model on
site $i$, which can take three different values, $o,\uparrow$ and
$\downarrow$, representing the hole, spin-up electron and spin-down
electron states.   We choose the hole representation in our
calculations, and thus the hole state has an odd parity $P^f(o)=1$
while the electron states have even parity
$P^f(\uparrow,\downarrow)=0$. On each site, the non-zero components
of the Grassmann tensors should satisfy the parity conservation
constraint:
\begin{align}
P^f(m_{i})+P^f(a)+P^f(b)+P^f(c)=0 ({\rm mod}\, 2) .
\end{align}
Since the wavefunction Eq.(\ref{GTPS}) does not have a definite
fermion number, we use the grand canonical ensemble, adding a
chemical potential term to Eq.(\ref{tJ}) to control the average hole
concentration.

We use the imaginary time evolution method\cite{Gumethod} to update
the GTPS from a random state. Then we use the weighted
Grassmann-tensor-entanglement renormalization group (wGTERG)
method\cite{GuGTPS,Gumethod}
to calculate physical quantities.(See appendix sections for details.)
The total system size ranges up to $2\times 27^2$ sites and all
calculations are performed with periodic boundary conditions. The
largest virtual dimension of the GTPS considered is $12$. To ensure
convergence of the wTERG method, we keep $D_{cut}$ (defined in
Refs.~\cite{GuGTPS,Gumethod}) up to $130$ for $D=4,6,8,10$, which
gives relative errors for physical quantities of order $10^{-3}$. For
$D=12$, we keep $D_{cut}$ up to $152$.

\begin{figure}[t]
{\includegraphics[width=0.5\textwidth]{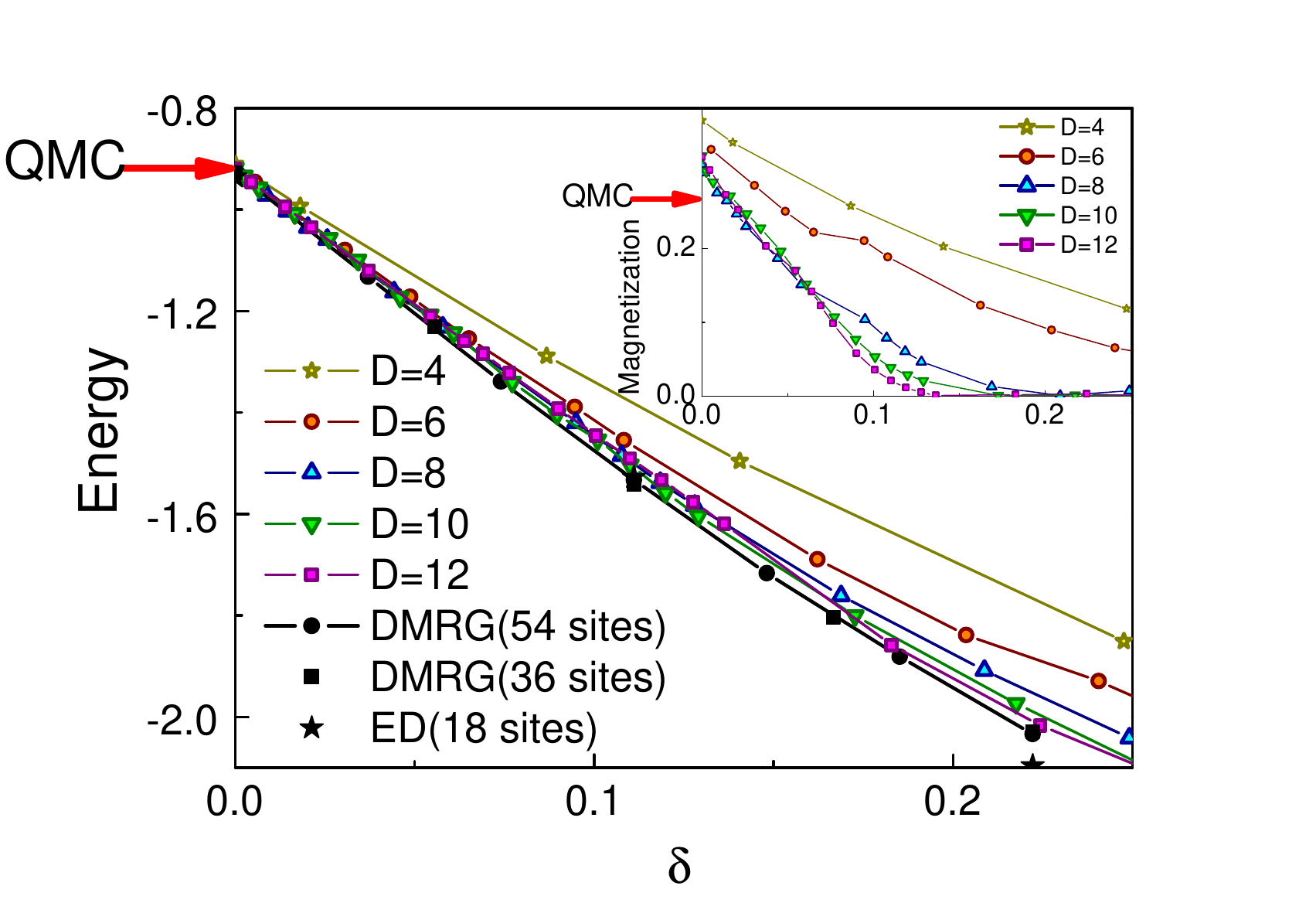}} \caption{(Color online)Ground
state energy as a function of doping. As a benchmark, we performed
the ED calculation and DMRG calculations for small system size.
Inset: Stagger magnetization as a function of
doping.}\label{fig:energy}
\end{figure}

\section{Ground state energy and staggered magnetization} At
half-filling, the $t$-$J$ model reduces to the Heisenberg model. In
this case, we find that the converged ground state energies per site are
$-0.5439$ for $D=10$ and $-0.5441$ for $D=12$(the term
$-\frac{1}{4}n_in_j$ is subtracted here), which are consistent with a
previous TPS study\cite{XiangTRG3} (with virtual dimension $D=5$ and
$6$, since all the components of GTPS with odd fermion parity
virtual indices vanish in this case) and a recent quantum Monte
Carlo (QMC) result $E=-0.54455(20)$.  Despite the good agreement with
the ground state energy, the staggered magnetization
$m=\sqrt{{\langle S^x_{i} \rangle}^2+{\langle S^y_{i} \rangle}^2+{\langle S^z_{i} \rangle}^2}$
(with $\v S_{i\in A}=-\v S_{i\in B}$ observed numerically) obtained
from our calculations is larger than the QMC result $m=0.2681(8)$.
We find $m=0.3257$ for $D=10$ and $m=0.3239$ for $D=12$, which are also
consistent with the previous TPS
study\cite{XiangTRG1,XiangTRG2,XiangTRG3}.

Nevertheless, we emphasize that the variational approach indeed does
obtain the correct phase. Actually, a recent study for square
lattice Heisenberg model shows that $m$ can be consistent with the
QMC result if $D$ is sufficiently
large\cite{LingTPS}.

Much more interesting physics arises after we dope the system.(We
consider $t/J=3$). As seen in Fig.\ref{fig:energy}, the ground state
energy shows a marked increase in $D$ dependence as
hole doping $\delta$ increases. As a benchmark, we perform the ED and DMRG
calculations for small periodic clusters with $N$ sites ($N=18$ for
ED and $N=36,54$ for DMRG).  These two methods are the only unbiased
methods for frustrated systems that avoid the sign problem, but they are
restricted to relatively small systems. To ensure the convergence of
the DMRG, we keep up to $8000$ states and make the truncation errors
less than $10^{-9}$ in our $N=54$ calculations. Up to $D=10$, we
find a systematic convergence of the ground energy.  Some data
points around $\delta=0.1$ for $D=12$ have slightly higher energy
than $D=10$, because $D_{cut}=152$ is still not large enough for
convergence at $D=12$.

As shown in the insert of Fig.\ref{fig:energy}, the staggered
magnetization $m$ has an even larger $D$ dependence than energy at
finite doping. However, up to $D=12$, the data appears to converge to a
relatively well-defined curve indicating vanishing AF order for $\delta
\gtrsim 0.1$. We also observed $\langle n_{i \in A} \rangle=\langle n_{j\in B} \rangle$ for arbitrary $\delta$
and therefore there is no commensurate charge density wave(CDW) order.

\begin{figure}[t]
{\includegraphics[width=0.5\textwidth]{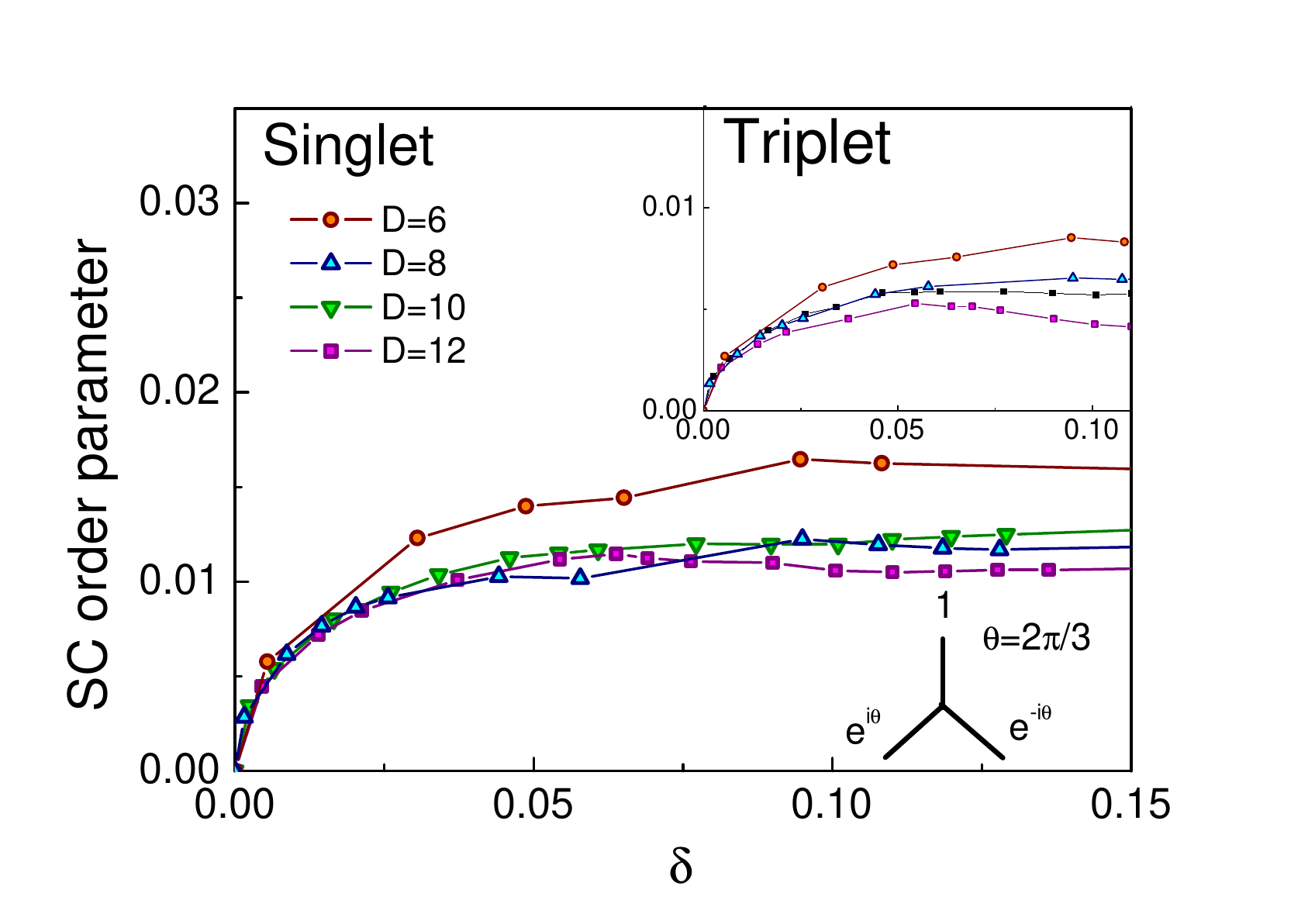}} \caption{ SC order
parameters as a function of doping.}\label{fig:SC}
\end{figure}

\section{Superconductivity}
 Next we turn to the interesting
question of whether the doped antiferromagnetic Mott insulator on
the honeycomb lattice supports superconductivity or not, and if so, what
its pairing symmetry is. To answer this, we calculate the real space
superconducting (SC) order parameters in the spin singlet channel
$\Delta^s=\frac{1}{\sqrt{2}}\left(c_{i,\uparrow}c_{j,\downarrow}-
  c_{i,\downarrow}c_{j,\uparrow}\right)$,
where $i$ and $j$ are nearest neighbor sites.Because we use a
chemical potential to control the hole concentration, the charge
$U(1)$ symmetry can be spontaneously broken in the variational
approach, which allows $\Delta^s$ to be measured directly rather
than through its two-point correlation function. As shown in the
main panel of Fig.\ref{fig:SC}, up to $\delta=0.15$ we find a
non-zero singlet SC order parameter for the whole region.
Strikingly, we find that the SC state breaks time reversal symmetry.
By measuring the SC order parameters for the three inequivalent
nearest-neighbor bonds, we found
$\Delta^s_a/\Delta^s_b\simeq\Delta^s_b/\Delta^s_c\simeq\Delta^s_c/\Delta^s_a\simeq e^{i\theta}$
with $\theta=\pm\frac{2\pi}{3}$(see Table \ref{phase}).  This pairing symmetry is usually
called $d+id$ wave.  We note that the above result is quite nontrivial since we start with a
completely random state without any pre-assuming SC order. Indeed, we observed that the emergence of such a $d+id$ wave SC
is a consequence of gaining kinetic energy(the $t$ term) of the $t$-$J$ model during the imaginary time evolution.
To exclude the possibility that this results
from trapping in an unstable local minimum, we repeat the
calculations with many different random tensors, and all cases
converge to the same results. Moreover, we also check that the SC
order parameter vanishes at large $t/J$ to make sure that the
existing SC order is the consequence of spontaneous symmetry
breaking.(For $D=6$, the critical value is around $15$ at
$\delta\sim0.3$ )

\begin{table}[h]
\begin{tabular}{|c||c|c|c|}
\hline
Doping & $\delta=0.034$ & $\delta=0.101$ & $\delta=0.129$  \\
\hline
$\Delta^s_a/\Delta^s_b$ & (-0.500,-0.866)   &  (-0.499,-0.867) & (-0.502,-0.863) \\
$\Delta^s_b/\Delta^s_c$ & (-0.500,-0.866)   &  (-0.501,-0.866) & (-0.498,-0.869) \\
$\Delta^s_c/\Delta^s_a$ & (-0.500,-0.866)   &  (-0.500,-0.866) & (-0.500,-0.866) \\
\hline
\end{tabular}
\caption{Up to a very high precision, we observed $\Delta^s_a/\Delta^s_b\simeq\Delta^s_b/\Delta^s_c\simeq\Delta^s_c/\Delta^s_a\simeq e^{-\frac{2\pi i}{3}}=(-\frac{1}{2},-\frac{\sqrt{3}}{2})$ at different hole doping for a GTPS ansatz(Here we use the data with inner dimension $D=10$ as a simple example.)}\label{phase}
\end{table}
The existence of SC order is observed in our numerical study up to
$\delta=0.4$. However, a much larger inner dimension $D$ is required
for the convergence of ground state energy at larger hole
concentration, which is beyond the scope of this paper.(The GTPS
variational ansatz we use in this paper is designed to help us
understand the nature of Mott physics; at large doping, the Mott
physics becomes less important and can be studied much better by
other methods.)

\section{Coexisting phases at low doping}
Interestingly, we find
the SC and AF order coexisting in the regime $0<\delta<0.1$. A
physical consequence of the microscopic coexistence is that triplet
pairing is induced.  The inset of Fig.\ref{fig:SC} shows the
amplitude of the triplet order parameter as a function of doping.
Since the triplet pairing order parameter has three independent
components $\vec\Delta_t = \frac{1}{\sqrt{2}}c_{i,\alpha} (i\sigma^y
\vec\sigma)_{\alpha\beta}
  c_{j,\beta} = {\bf d} e^{i\phi}$(Here $i \in A$, $j \in B$ and $\phi$ is the
phase of SC order parameter.), we can define the amplitude of
triplet order parameter as $\Delta_t = \sqrt{\vec\Delta_t^* \cdot
\vec\Delta_t}$.
The phase shift of $2\pi/3$ on the three inequivalent bonds is also
observed for all the triplet components. We further check the
internal spin direction of the triplet $\v d$ vector and find that it is
always anti-parallel to the Neel vector($\v S_{\rm{Neel}}= \langle
\v S_i\rangle-\langle \v S_j \rangle$). At larger doping
$\delta>0.1$, the triplet order parameter has a very strong $D$
dependence, so at present we are unable to determine whether it
ultimately vanishes or remains non-zero in the $D\rightarrow \infty$
limit. We leave this issue for future work. By fully using all
symmetry quantum numbers and other techniques like high performance
simulation on GPUs, we can in principle deal with $D$ up to $20-30$.

\section{Discussions}
The variational ansatz in this paper ignores any possible
incommensurate phases. Natural candidates are
spiral or striped anti-ferromagnetic phases, as having been intensively
discussed for the square lattice. A weak coupling perspective
suggests that this may be unlikely.

In the weak-coupling limit of the Hubbard model on honeycomb
lattice, the system is a semimetal with two Dirac cones. With a
non-zero Hubbard $U$, commensurate AF fluctuations at zero momentum
manifest in the inter-band susceptibility, becoming stronger and
stronger with increasing $U$. At sufficiently large U commensurate
AF order develops (recent numerics find a narrow region of
intermediate spin liquid phase\cite{honeycombSL}).  At small but
finite doping, the Dirac cones become pockets. In this case, the
total intra-band spin susceptibility shows a constant behavior for
small $q$($q<2k_f$)\cite{graphene} while the total inter-band spin
susceptibility still has a peak at zero momentum. Thus, we argue that the
commensurate AF fluctuations at zero momentum still dominate for
sufficiently small hole concentration.

\begin{figure}[tbp]
\begin{center}
\vskip 0.0cm \hspace*{-0.0cm}
\subfigure[]{\includegraphics[width=1.0\columnwidth]{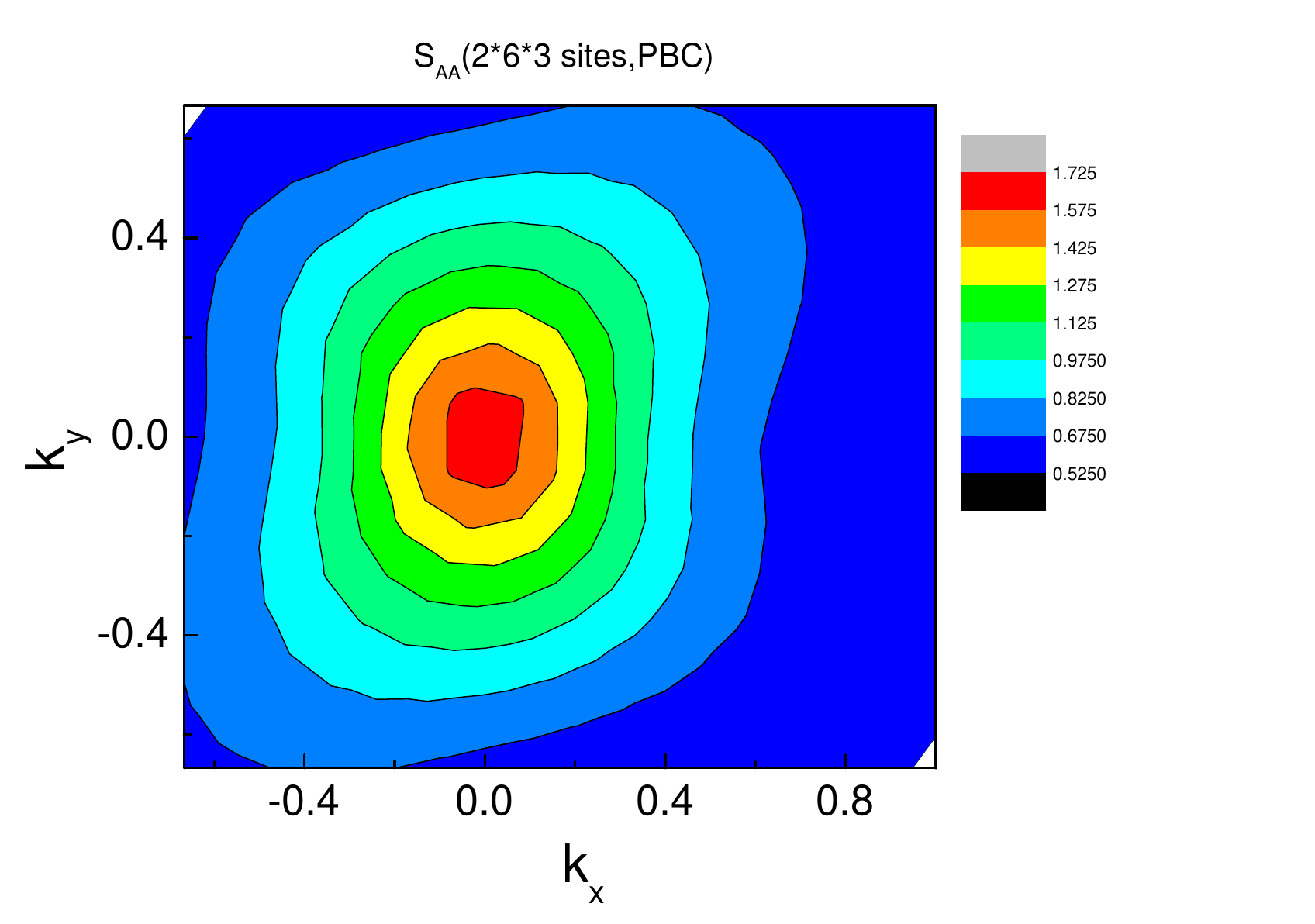}\label{fig:spinAA}}\\
\subfigure[]{\includegraphics[width=1.0\columnwidth]{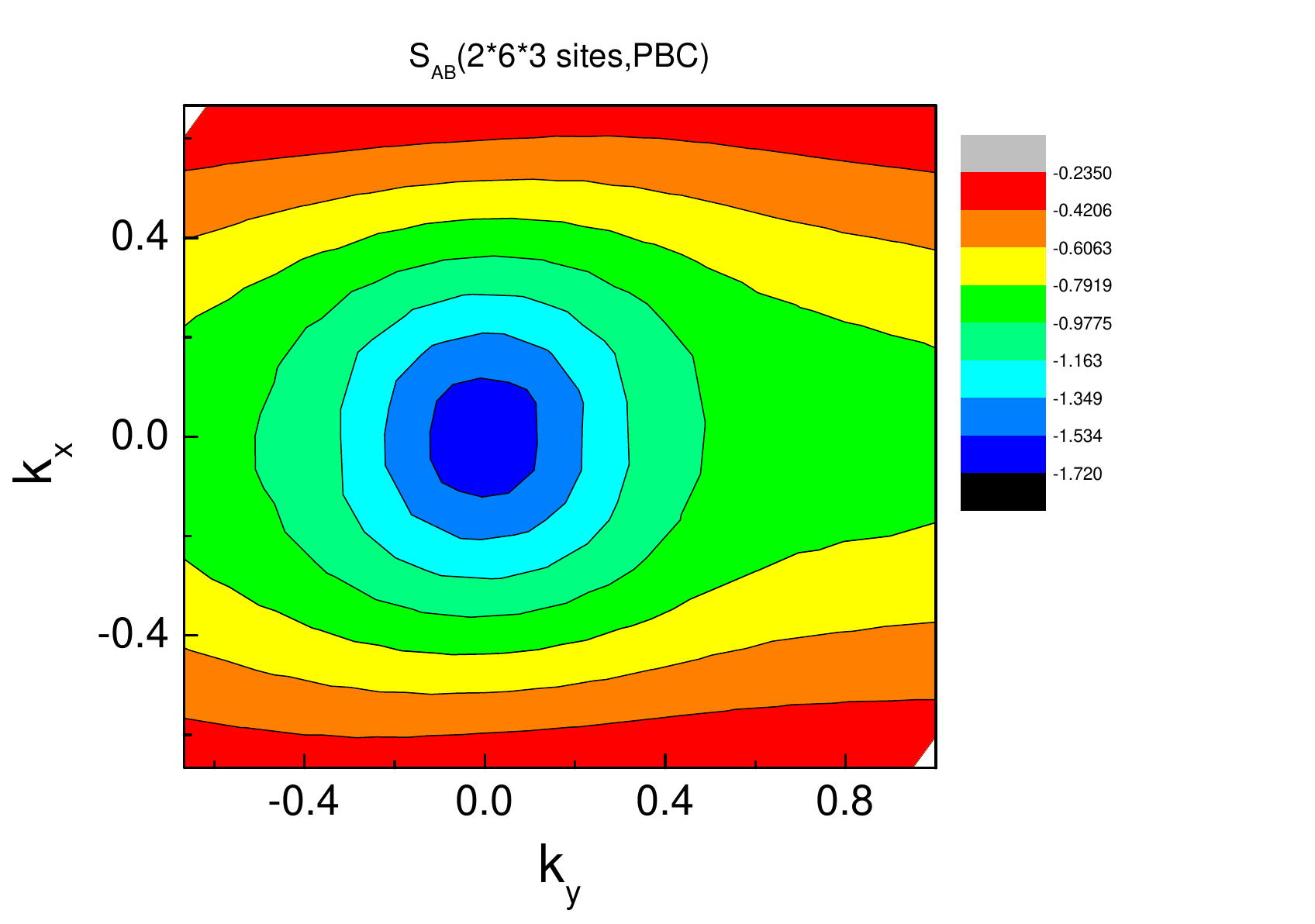}\label{fig:spinAB}}
\caption{(Color online) Counter plots for the spin structure factors for the same sublattice and for different sublattices. We obtain the results by performing high precision DMRG algorithm on a small cluster($2*6*3$ sites) at $5.5\%$ doping with periodic boundary condition(PBC).}
\end{center}
\end{figure}

Our DMRG calculations also support this argument. Up to 54
sites, we do not find any evidence for incommensurate spin-spin
correlation.
As seen in Figs. \ref{fig:spinAA} and \ref{fig:spinAB}, we plot the spin structure factors for the same sublattice and for different sublattices:
\begin{align}
S_{AA}(\v k)&=\frac{1}{N}\sum_{i\in A;j \in A}e^{i\v k \cdot (\v r_i-\v r_j)} \langle \v S_i \cdot\v S_j\rangle\nonumber\\
S_{AB}(\v k)&=\frac{1}{N}\sum_{i\in A;j \in B}e^{i\v k \cdot (\v r_i-\v r_j)} \langle \v S_i \cdot\v S_j\rangle,
\end{align}
where $N$ is the total number of unit cells, which is $18$($6*3$) in our DMRG calculation. We find that both $S_{AA}$ and $S_{AB}$ have peaks at $\v k=(0,0)$, with a positive value and a negative value. Such a result implies a ferromagnetic long range order for the same sublattice but a anti-ferromagnetic long range order for different sublattices. Although the system size in our DMRG calculation may be not large enough, we believe the conclusion is still correct in the thermodynamic limit since other calculation, e.g., meanfield theory, also supports this result. On the other hand, as we know that superconductivity is incomparable with incommensurate magnetic orders, any magnetic order that coexists with superconductivity must be commensurate.

\begin{figure}[t]
{\includegraphics[width=0.4\textwidth]{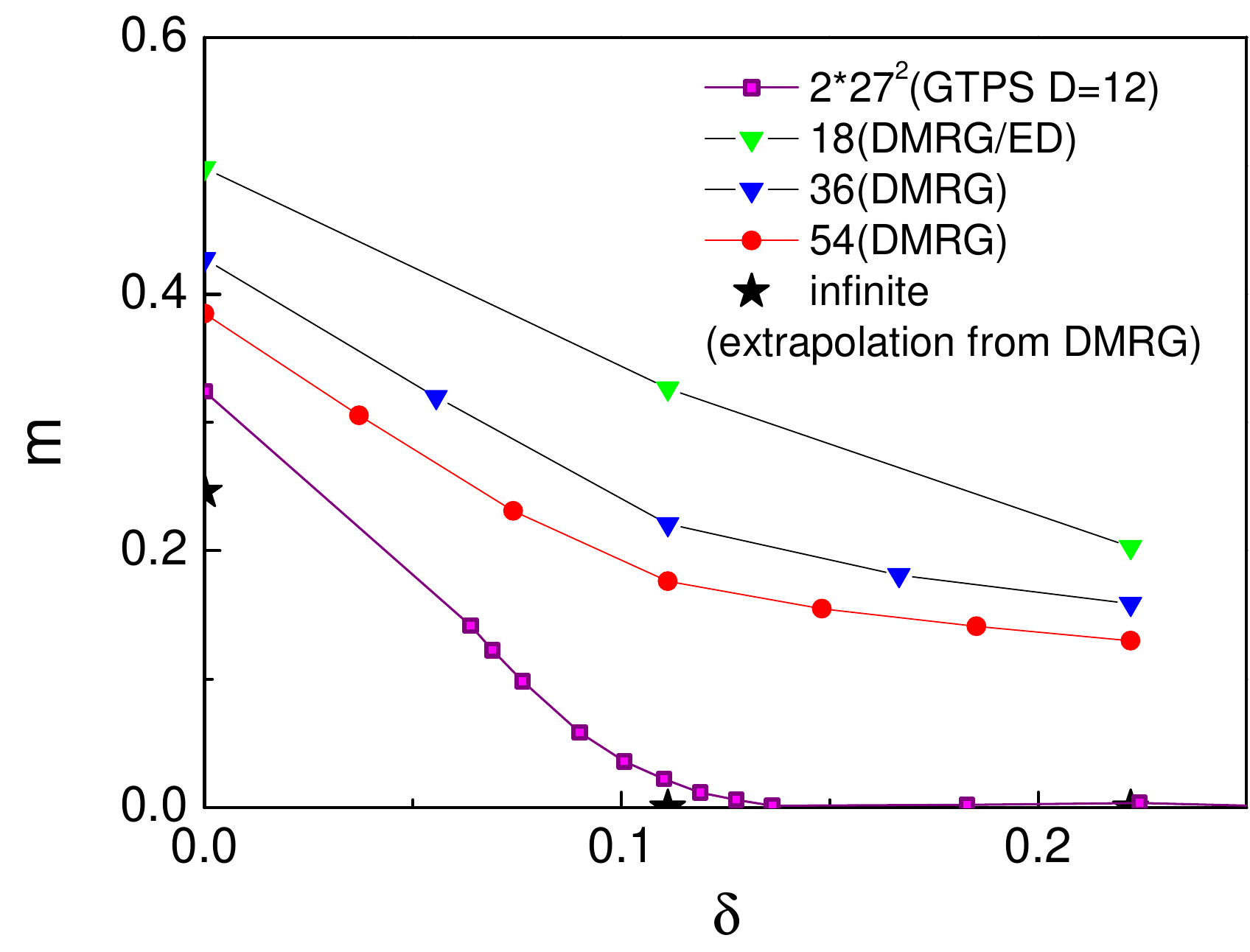}} \caption{(Color online)The
comparison of DMRG/ED and GTPS calculations for stagger
magnetization $m$. }\label{fig:compare}
\end{figure}

On the other hand, we find that the GTPS results are comparable with
extrapolations of (commensurate)staggered magnetization $m$ for infinite size systems(the GTPS results
for $m$ are somewhat larger when close to half-filling due to
insufficient tensor dimension $D$). Fig. \ref{fig:compare} shows the comparison of DMRG
calculations and GTPS calculations for staggered magnetization $m$
in small systems.

A more general concern is whether the GTPS tends to overestimate SC
order at large doping, due to its non-conservation of charge(note
that the projected wavefunction approach has a rather strong
tendency to produce superconducting states). The observed SC order
parameter is ``small'' in terms of the natural upper limit $\langle
cc\rangle \lesssim 0.1\delta \ll \delta$.  Nevertheless, our prediction of $d+id$ paring symmetry
is supported by other approaches: (a) in the mean field theory
for the honeycomb $t$-$J$ model, only the $d+id$ pairing channel gains
energy\cite{tJMF,tJVQMC}, therefore $d+id$ pairing symmetry is most possible if the
ground state of $t$-$J$ model is a superconductor and (b) in the weak coupling limit of the
Hubbard model, very recent renormalization group studies also find
$d+id$ superconductivity around quarter
filling\cite{weak1,weak2,RGhubbard1,RGhubbard2,ronny}. However, we believe that the mechanism
of superconductivity discovered at low doping is a consequence of strong interaction and is quite different from the weak coupling case.
Apparently our results can not be explained by any weak coupling theory, as the Dirac cone is stable at low doping(e.g., $\delta<0.1$) and there is no superconductivity. The recently
proposed skyrmion superconductivity is a very promising candidate\cite{skyrmionSC} and we will
explore this kind of idea in our future work.

In conclusion, we report the theoretical discovery of a $d+id$ wave
superconducting ground state in the $t$-$J$ model on a honeycomb lattice, based
on a recently developed variational method - the GTPS approach. At
low doping, AF order coexists with the SC order. In the
coexistence regime, a spin triplet pairing with the same phase shift
is induced and its triplet $\textbf{d}$ vector is anti-parallel with
the Neel vector. It would be interesting to search for this physics
in experiment. The recently discovered spin $1/2$ honeycomb lattice
antiferromagnet InV$_{1/3}$Cu$_{2/3}$O$_3$\cite{InVCuO} would be an
appealing candidate if it could be doped experimentally.

\section*{Acknowledgment}
The authors would like to thank F. Verstraete, J. I. Cirac, M. P. A.
Fisher, F. C. Zhang, P. A. Lee, Z. Y. Weng, L. Fu, K. Sun, C. Wu, F.
Yang and Y. Zhou for valuable discussions. Z.C.G. is supported by
NSF Grant No. PHY05-51164; H.C.J was supported in part by the
National Basic Research Program of China Grant 2011CBA00300,
2011CBA00302; D.N.S is supported by DMR-1205734 and DMR-0906816 ;
L.B. is supported by NSF grant DMR-0804564 and a Packard Fellowship;
H.Y. was partly supported by DOE grant DE-AC02-05CH11231; X.G.W. is
supported by NSF Grant No. DMR-1005541 and NSFC 11074140.

\begin{appendix}
\section{The imaginary time evolution algorithm of GTPS}
In this paper, we use the (simplified) imaginary time evolution
method\cite{Gumethod} to update the GTPS variational wave function.
In the hole representation, we can decompose the
$\tilde{c}^\dagger_{i,\sigma}$ as $\tilde{c}_{i,\sigma}=h^\dagger_i
b_{i;\sigma}$. Here the holon $h_i$ is a fermion while the spinon
$b_{i;\sigma}$ is a boson. The no-double-occupancy constraint reads:
\begin{align}
\sum_{\sigma} b_{i;\sigma}^\dagger b_{i;\sigma}+h_i^\dagger
h_i=1\label{constraint}
\end{align}
Under this representation, we can rewrite the $t-J$ model as:
\begin{equation}
H_{t-J}=-t\sum_{\langle ij\rangle,\sigma }h^\dagger_jh_i
b^\dagger_{i;\sigma}b_{j;\sigma}+h.c.+J\sum_{\langle ij
\rangle}\left(\v S_i \v S_j-\frac{1}{4}n_i^b n_j^b\right),\label{tJ2}
\end{equation}
where
\begin{align}
\v S_i=\sum_{\sigma\sigma^\prime}b_{i;\sigma}^\dagger
\mathbf{\tau}_{\sigma\sigma^\prime} b_{i;\sigma^\prime};\quad
n^b_i=\sum_\sigma b^\dagger_{i;\sigma}b_{i;\sigma}
\end{align}
Due to the no-double-occupancy constraint Eq. (\ref{constraint}),
the spin up/down states
$|\uparrow(\downarrow)_i\rangle=b_{i;\uparrow(\downarrow)}^\dagger|0\rangle$
and the hole state $|o\rangle=h_i^\dagger|0\rangle$ form a complete
basis for each site. The closure condition reads:
\begin{align}
|\uparrow_i\rangle \langle \uparrow_i|+|\downarrow_i\rangle \langle
\downarrow_i|+|o_i\rangle \langle o_i|=1\label{closure}
\end{align}

As already has been discussed in Ref.\cite{Gumethod}, we need to use
the fermion coherent state representation to perform the imaginary
time evolution algorithm for GTPS. Let us introduce the fermion
coherent state of holon $|\eta_i\rangle=|0\rangle-\eta_i
h_i^\dagger|0\rangle$.($\eta_i$ is a Grassmann variable here.) In
this new basis, the closure relation Eq. (\ref{closure}) becomes:
\begin{align}
|\uparrow_i\rangle \langle \uparrow_i|+|\downarrow_i\rangle \langle
\downarrow_i|+\int \dd \bar\eta_i \dd \eta_i|\eta_i\rangle \langle
\bar\eta_i|=1\label{closure}
\end{align}
The variational ground state can be determined through imaginary
time evolution:
\begin{align}
|\Psi_G\rangle=e^{-\tau H_{t-J}}|\Psi_0\rangle, \quad \tau
\rightarrow\infty
\end{align}
For a sufficiently thin time slice $\delta\tau$, we can decompose
$e^{-\delta\tau H_{t-J}}$ as:
\begin{align}
e^{-\delta\tau H_{t-J}}\sim e^{-\delta\tau H_{t-J}^x}e^{-\delta\tau
H_{t-J}^y}e^{-\delta\tau H_{t-J}^z}
\end{align}
Here $x,y,z$ represent three different directions of the honeycomb
lattice and each $H_{t-J}^{x(y,z)}$ term contains a summation of
non-overlapped two body Hamiltonian $H_{t-J}^{x(y,z)}=\sum_{\langle
ij \rangle}h_{ij}^{x(y,z)}$. Thus, for a sufficiently thin time slice,
we can apply the evolution operator along the $x(y,z)$ direction
separately.

%In the fermion coherent state representation, the two body evolution
%operator along $x$ direction $e^{-\delta h_{ij}^x}$ can be expressed
%as a Grassmann valued matrix acting on the two sites Hilbert space:
%\begin{align}
%&\langle \mathfrak{m}_i \mathfrak{m}_j |e^{-\delta\tau
%h_{ij}^x}|\mathfrak{m}_i^\prime \mathfrak{m}_j^\prime\rangle
%\nonumber\\=&E_{m_i^\prime m_j^\prime}^{m_i
%m_j}(\eta_j)^{P^f(m_j)}(\eta_i)^{P^f(m_i)}{(\bar\eta_i^\prime)}^{P^f(m_i^\prime)}{(\bar\eta_j^\prime)}^{P^f(m_j^\prime)}
%\end{align}
%Here $\mathfrak{m}_i$ is the fermion coherent state indices valued
%as $\uparrow$, $\downarrow$ or $\eta$ while $m_i$ is the usual Fock
%state indices valued as $\uparrow$, $\downarrow$ or $o$. We notice
%the spin states remain the same in both representations.

\begin{figure}[h]
{\includegraphics[width=0.5\textwidth]{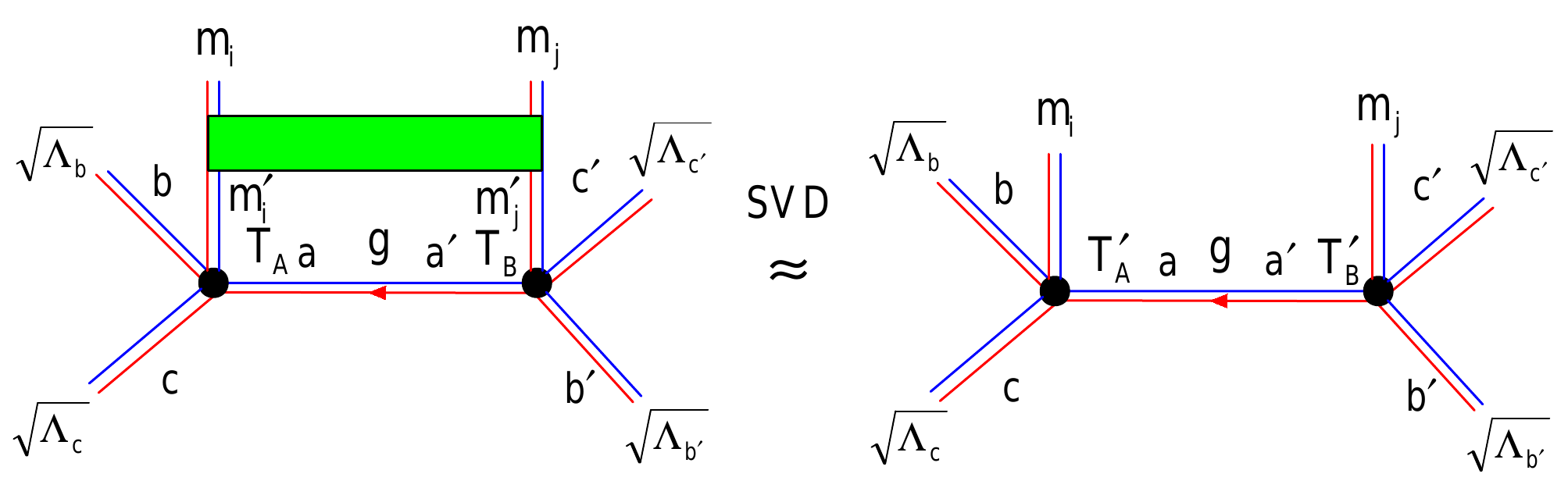}} \caption{(Color online)A
schematic plot for the imaginary time evolution algorithm.
}\label{fig:evolution}
\end{figure}

\begin{figure}[h]
{\includegraphics[width=0.5\textwidth]{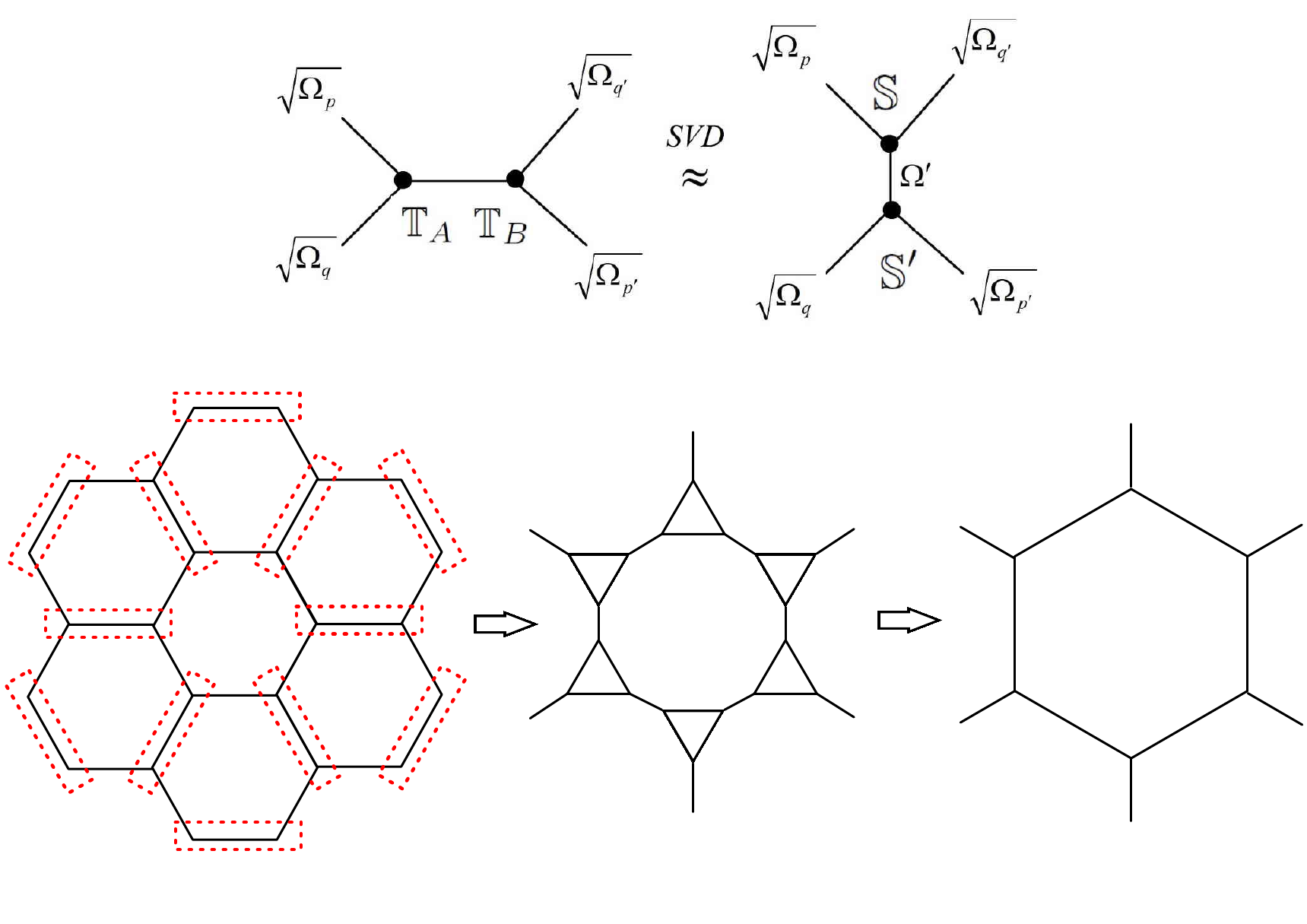}} \caption{(Color online)A
schematic plot for the renormalization algorithm on honeycomb
lattice. Similar to the simplified imaginary time evolution
algorithm, we use a weighting vector $\sqrt{\Omega}$ to mimic the
environment effect. As has been discussed in Ref.\cite{Gumethod},
the initial value of $\Omega$ can be determined by $\Lambda$ on the
corresponding link and will be updated during the RG
scheme.}\label{fig:honeycombRG}
\end{figure}

By applying the method developed in Ref.\cite{Gumethod}, we can
successfully update the (complex) variational parameters
${T}^{m_{i}}_{A;abc}$ and ${T}^{m_{j}}_{B;a^\prime b^\prime
c^\prime}$ as:
\begin{align}
{{T}}^{\prime
m_i}_{A;abc}&=(-)^{P^f(m_i)P^f(a)}\frac{\sqrt{\Lambda_a^\prime}}{\sqrt{\Lambda_b^y}
\sqrt{\Lambda_c^z}}U_{bcm_i;a}  \nonumber\\
{{T}}^{\prime m_j}_{B;a^\prime b^\prime c^\prime
}&=(-)^{P^f(m_j)P^f(a^\prime)}\frac{\sqrt{\Lambda_{a^\prime}^\prime}}
{\sqrt{\Lambda_{b^\prime}^y}\sqrt{\Lambda_{c^\prime}^z}}V_{b^\prime
c^\prime m_j;a^\prime},\label{newT}
\end{align}
where $U$ and $V$ are determined by the singular value
decomposition(SVD) of the following matrix $M$:
\begin{align}
&{\rm{M}}_{bcm_i;b^\prime c^\prime m_j}  = \sum_{am_i^\prime
m_j^\prime}\sqrt{\Lambda_b^y}
\sqrt{\Lambda_c^z}\sqrt{\Lambda_{b^\prime}^y}\sqrt{\Lambda_{c^\prime}^z}\nonumber\\&\times
(-)^{\left[P^f(m_i^\prime)+P^f(m_j^\prime)\right]P^f(a)}
(-)^{P^f(m_j^\prime)\left[P^f(m_i^\prime)+P^f(b)+P^f(c)\right]} \nonumber\\
&\times (-)^{m_j\left[P^f(m_i)+P^f(b)+P^f(c)\right]}  E_{m_i^\prime
m_j^\prime}^{m_i m_j}{
T}^{m_{i}^\prime}_{A;abc}{T}^{m_{j}^\prime}_{B;a b^\prime
c^\prime}\label{M}
\end{align}
Here $E_{m_i^\prime m_j^\prime}^{m_i m_j}$ is the matrix elements of
the evolution operator $e^{-\delta h_{ij}^x}$ under the Fock basis. We
keep the largest $D$th singular values:
\begin{align}
{\rm{M}}_{bcm_i;b^\prime c^\prime m_j}\simeq\sum_{a=1}^D
U_{bcm_i;a}\Lambda_a^\prime V_{b^\prime c^\prime m_j;a}
\end{align}

Similar to the usual TPS case\cite{XiangTRG1}, the environment
weight vectors $\Lambda^{x(y,z)}$ can be initialized as $1$ and then
updated during the time evolution. For example, $\Lambda^x$ is
updated as $\Lambda^\prime$ in the above evolution scheme.

\begin{figure}[h]
{\includegraphics[width=0.5\textwidth]{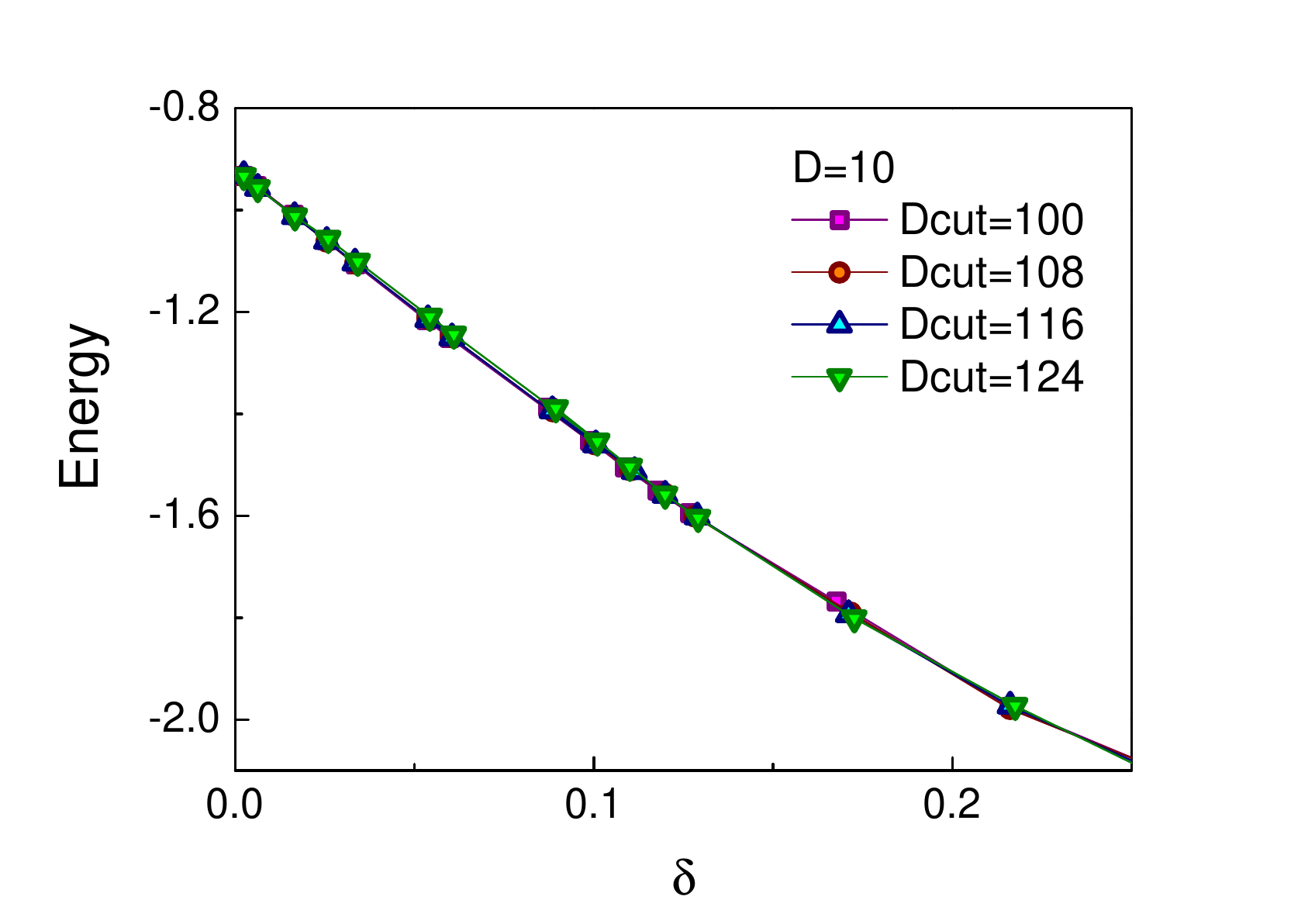}}
\caption{(Color online)Variational ground state energy($D=10$) as a function of
doping with different $D_{cut}$(Number of eigenvalues kept in the
wGTERG algorithm). It is shown that all the data points almost collapse
on the same curve. We find that the relative error is of order $10^{-3}$.
}\label{fig:error}
\end{figure}

\section{The GTERG/wGTERG algorithm}
After determining the
variational ground state of GTPS by performing the imaginary time
evolution, we can compute the physical quantities by using the
GTERG/wGTERG method developed in Refs\cite{GuGTPS,Gumethod}. This
method can be regarded as the Grassmann variable generalization of
the usual TERG/wTERG method\cite{GuTERG,Gumethod}, where a coarse
graining procedure is designed to calculate the physical quantities
of TPS efficiently. In Fig.\ref{fig:error}, we plot the variational
ground state energy($D=10$) as a function of doping with different
$D_{cut}$(Number of eigenvalues kept in the wGTERG algorithm). We
find that the relative error is of order $10^{-3}$.

\end{appendix}

\end{document}